\documentclass[sigconf]{acmart}
\AtBeginDocument{%
  \providecommand\BibTeX{{%
    \normalfont B\kern-0.5em{\scshape i\kern-0.25em b}\kern-0.8em\TeX}}}

\setcopyright{acmcopyright}
\copyrightyear{2018}
\acmYear{2018}
\acmDOI{10.1145/1122445.1122456}

\acmConference[Woodstock '18]{Woodstock '18: ACM Symposium on Neural
  Gaze Detection}{June 03--05, 2018}{Woodstock, NY}
\acmBooktitle{Woodstock '18: ACM Symposium on Neural Gaze Detection,
  June 03--05, 2018, Woodstock, NY}
\acmPrice{15.00}
\acmISBN{978-1-4503-XXXX-X/18/06}

\usepackage{lineno}
\usepackage{subfigure}
\usepackage{bm}
\usepackage{amsmath}
\usepackage{multirow}
\usepackage{float}
\usepackage{colortbl}
\usepackage{hyperref}
\usepackage{enumitem}
\usepackage[ruled]{algorithm2e}

\newtheorem{Framework}{\textbf{Framework}}

\newtheorem{Definition}{\textbf{Definition}}

\newcommand{\squishlist}{
   \begin{list}{$\bullet$}
    { \setlength{\itemsep}{0pt}      \setlength{\parsep}{2pt}
      \setlength{\topsep}{2pt}       \setlength{\partopsep}{0pt}
      \setlength{\leftmargin}{1em}   \setlength{\labelwidth}{1em}
      \setlength{\labelsep}{0.5em}   \setlength{\itemindent}{0pt}
      \setlength{\listparindent}{0pt}} }

\newcommand{\squishend}{
    \end{list}  }
    
\newcommand{\topcaption}{%
\setlength{\abovecaptionskip}{3pt}%
\setlength{\belowcaptionskip}{0pt}%
\caption}

\def\firstmodelname{HCB}
\def\secondmodelname{pHCB}

\DeclareMathOperator*{\argmax}{arg\,max}

\begin{document}
 
\title{Show Me the Whole World: Towards Entire Item Space Exploration for Interactive Personalized Recommendations}

\author{Yu Song}
\email{yusonghust@gmail.com}
\affiliation{
    \institution{Huazhong University of Science and Technology}
    \city{Wuhan}
    \country{China}
}

\author{Jianxun Lian}
\email{jianxun.lian@outlook.com}
\affiliation{
    \institution{Microsoft Research Asia}
    \city{Beijing}
    \country{China}
}

\author{Shuai Sun}
\email{540507710@qq.com}
\affiliation{
    \institution{Huazhong University of Science and Technology}
    \city{Wuhan}
    \country{China}
}

\author{Hong Huang}
\email{honghuang@hust.edu.cn}
\affiliation{
    \institution{Huazhong University of Science and Technology}
    \city{Wuhan}
    \country{China}
}

\author{Yu Li}
\email{liyu65@meituan.com}
\affiliation{
    \institution{Meituan Group}
    \city{Beijing}
    \country{China}
}

\author{Hai Jin}
\email{hjin@hust.edu.cn}
\affiliation{
    \institution{Huazhong University of Science and Technology}
    \city{Wuhan}
    \country{China}
}
  
\author{Xing Xie}
\email{xingx@microsoft.com}
\affiliation{
    \institution{Microsoft Research Asia}
    \city{Beijing}
    \country{China}
}


\begin{abstract}

User interest exploration is an important and challenging topic in recommender systems, which alleviates the closed-loop effects between recommendation models and user-item interactions. Contextual bandit (CB) algorithms strive to make a good trade-off between exploration and exploitation so that users’ potential interests have chances to expose. However, classical CB algorithms can only be applied to a small, sampled item set (usually hundreds), which forces the typical applications in recommender systems limited to candidate post-ranking, homepage top item ranking, ad creative selection, or online model selection (A/B test). 

In this paper, we introduce two simple but effective hierarchical CB algorithms to make a classical CB model (such as LinUCB and Thompson Sampling) capable to explore users’ interest in the entire item space without limiting to a small item set. We first construct a hierarchy item tree via a bottom-up clustering algorithm to organize items in a coarse-to-fine manner. Then we propose a hierarchical CB (HCB) algorithm to explore users’ interest on the hierarchy tree. HCB takes the exploration problem as a series of decision-making processes, where the goal is to find a path from the root to a leaf node, and the feedback will be back-propagated to all the nodes in the path. We further propose a progressive hierarchical CB (pHCB) algorithm, which progressively extends visible nodes which reach a confidence level for exploration, to avoid misleading actions on upper-level nodes in the sequential decision-making process. Extensive experiments on two public recommendation datasets demonstrate the effectiveness and flexibility of our methods.

\end{abstract}


\maketitle


\section{Introduction}
Recommender systems help users to easily find their favorite items from massive candidates. Typically, recommender models, such as collaborative filtering~\cite{koren2009matrix} and DeepFM~\cite{10.5555/3172077.3172127}, exploit users’ historical behaviors to learn users’ preference for future recommendations. Recommender systems with only exploitation models usually suffer from closed-loop effects~\cite{jadidinejad2020using}: users mostly only interact with the items recommended by the system; the system further consolidates users' profiles with their interacted items recommended by the deployed model. Therefore, as time goes on, the system will be biased to a small, exposed set of interests for each user and keep recommending a limited range of items to a same user. 

Contextual multi-armed bandit algorithms, such as LinUCB~\cite{li2010contextual}, are classical methods that leverage side information to provide a good trade-off between exploration and exploitation, so that the closed-loop effect can be alleviated. Items are treated as arms and the recommender model is treated as an agent. Basically, at each round, the agent chooses one arm which has the biggest potential from $K$ arm candidates, then receives a corresponding reward based on user-item interaction. The goal is to maximize the cumulative reward over $T$ rounds. However, these algorithms hold a premise that $K$ is small, so enumerating all arm candidates' scores and pick up the best one is feasible. The premise is true for a few scenarios where the candidates are naturally small, for example, homepage breaking news ranking, ads creative ranking and online model selection. In the scenario of general recommender systems, to fully explore users' potential interests and truly alleviate the closed-loop effect, the arms candidates are the entire item repository, which usually contains millions or even billions of items. Classical bandit methods become infeasible due to the high computation cost of enumerating every one of the arms. 


To address the challenge, we first propose a generic hierarchical contextual bandit~ (\textbf{\firstmodelname}) algorithm to efficiently explore the interests of users for large-scale recommendation scenarios. Tree structure are widely employed to partition the search space to reduce the computational cost~\cite{jain2016extreme,prabhu2014fastxml,zhu2018learning,zhuo2020learning}. {\firstmodelname} uses a tree structure as the index for coarse-to-fine retrieval. For example, in e-commerce scenario, \textsl{``Apparel > Clothing > Women's Clothing > Dresses''} is a path from general apparel to women's dresses. Instead of using the category taxonomy of items, we utilize a bottom-up clustering method on item embeddings to organize items as a hierarchy tree, on which each node contains a group of semantically similar items. As a result, the number of items associated to each node on the hierarchy tree can be balanced, and users' collaborative behaviors (such as co-click relations) can be encoded to form the hierarchy tree.
{\firstmodelname} leverages the hierarchical information and turns the interest exploration problem into a series of decision-making problems. Starting from the root of the tree, on each non-leaf node, {\firstmodelname} performs a bandit algorithm among the children arms to choose a child node until a leaf node is reached. Afterward, another bandit algorithm is responsible for recommending an item from the leaf node to the user and collect her feedback as reward. The reward will be back-propagated along the path to adjust the estimation of users' interest towards the hierarchy tree.

The process of {\firstmodelname} is like a depth-first search (DFS) idea. However, selecting a path in this DFS manner may cause new uncertainties, especially for a deep tree. First, if the selection of the parent node is misleading, all the subsequent choices will be impacted, which we call the \textsl{error propagation}. Second, since user interests are usually diverse, it is possible that the user is interested in many child nodes located in different parts of the tree. Therefore, we further propose a progressive {\firstmodelname} (\textbf{\secondmodelname}) algorithm to reduce uncertainties and enhance the capacity of recommendation. Like the process of breadth-first search (BFS), {\secondmodelname} explores items in an adaptive top-down manner. On the one hand, it gradually maintains a limited number of nodes as a receptive field. If one node has been explored multiple times and the user's interest on this node has been verified, the node's children nodes will be included to the receptive filed while the current node will be removed. On the other hand, {\secondmodelname} learns user interests of different aspects by performing a bandit algorithm with visible nodes in the receptive field as arms. Consequently, the {\secondmodelname} avoids greedily selecting only one node at each level to improve the {\firstmodelname}. To summarize, we make the following contributions:
\squishlist
\item We highlight the importance of exploring users' interests in the entire item space to truly alleviate the closed-loop effect in personalized recommender systems. To the best of our knowledge, it is the first attempt to implement CB models on millions of items. 
\item Two simple yet effective algorithms, i.e. {\firstmodelname} and {\secondmodelname},  are proposed to explore potential interests of users efficiently through a hierarchy item tree.
\item We conduct experiments on two large-scale recommendation datasets. Results show the superiority of {\firstmodelname} and {\secondmodelname} over baselines, as well as the flexibility to integrate with different exploration methods such as LinUCB, Thompson Sampling and $\epsilon$-greedy. In addition, we design an experiment to verify that thanks to the exploring mechanism, both {\firstmodelname} and {\secondmodelname} can effectively alleviate the closed-loop effects in recommender systems and learn better user profiles in the long term. 
\squishend


\section{Related Work}\label{related}
It is the first work to study entire space user interest exploration. Our work is relevant to two lines of research, and we will review them separately.

\subsection{Contextual Bandit Algorithms} 
Contextual bandit algorithms aims to seek a balance between exploration and exploitation, which have been used in several applications, such as recommender systems~\cite{mary2015bandits}, dynamic pricing~\cite{misra2019dynamic}, quantitative finance~\cite{shen2015portfolio} and so on. \cite{bouneffouf2019survey} reviews the existing practical applications of contextual bandit algorithms. By assuming the payoff model is linear, LinUCB~\cite{li2010contextual} and Thompson Sampling~\cite{agrawal2013thompson} and two representative methods for solving contextual bandit problems. Beyond them, a variety of algorithms have been proposed to optimize the performance or learning speed. For example, ConUCB~\cite{zhang2020conversational} introduces conversations between the agent and users to ask whether the user is interested in a certain topic occasionally. HATCH~\cite{yang2020hierarchical} considers the resource consumption of exploration and proposes a strategy to conduct bandit exploration with budget limitation. S-MAB~\cite{fouche2019scaling} considers two aspects, one is to maximize the cumulative rewards and the other is to decide how many arms to be pulled so as to reduce the exploration cost. GRC~\cite{wu2020learning} develops a graph regularized cross model to leverage the non-linearity of neural networks for better estimating the rewards. Different from them, our work commits to efficiently explore user interests in the entire space, rather than from a small subset of items.

\subsection{Cluster-of-Bandit Algorithms}
In the past few years, cluster-of-bandit algorithms have attracted the attention of some scholars. Generally, cluster-of-bandit algorithms aim to model the dependency since the items or users are always related to each other. As a result, cluster-of-bandit algorithms achieve better cumulative rewards than traditional contextual bandit algorithms due to knowledge sharing. For example, CLUB~\cite{gentile2014online}, DYNUCB~\cite{nguyen2014dynamic}, CAB~\cite{gentile2017context} and COFIBA~\cite{li2016collaborative} assign users with similar interests into a same subset to make decisions together, thus it make contributions to accelerate the learning speed. Different from them, this paper focus on modeling the item dependency. ICTRTS and ICTRUCB~\cite{wang2018online} explicitly model the item dependencies via clustering of arms, but they are only designed for context-free bandits. Similarly,  \cite{pandey2007bandits} uses a taxonomy structure to exploit arm dependencies with context-free bandits. Considering that context-free bandits cannot utilize the abundant side information for making decisions, their exploration ability has yet to be improved. HMAB~\cite{wang2018hierarchical} leverages a tree-structured hierarchy constructed by domain experts to design a hierarchical multi-armed bandit algorithm for online IT ticket automation recommendation. However, domain knowledge is hard to collect and HMAB can not be applied to large-scale recommender systems because it needs to traverse all the paths in the tree. Moreover, HMAB aims to learn latent parameters for the nodes in the hierarchy tree, which is totally different from our goal of exploring users' latent interests. Distributed bandit algorithms, such as DCCB~\cite{korda2016distributed} and DistCLUB~\cite{mahadik2020fast}, 
aim to speed up the computation by distributing the workloads in parallel. However, these methods do not address the issue of searching from tremendous items, the computational cost is still too expensive for responding users' requests in an online manner (for example, how to response 100 users' concurrent requests within 10 milliseconds in a scenario involving one million items). 
In summary, compared with existing cluster-of-bandit algorithms, our HCB and pHCB algorithms leverage a bottom-up clustering method to build a hierarchical tree of items, then explore users' potential interests in the entire space of items based on the item hierarchy.

\begin{table}[tbp]
\small
\topcaption{A collection of notations}
\setlength{\tabcolsep}{0.5mm}
\renewcommand\arraystretch{1.2}
\begin{tabular}{c|p{6.5cm}}
\hline
\textbf{Notation} & \ \ \ \ \ \ \ \ \ \ \ \ \ \ \ \ \ \ \ \ \ \textbf{Description} \\ \hline \hline
$\mathcal{U}$                 & User set, $\mathcal{U}$ = $\{u^{(1)}, u^{(2)}, \cdots, u^{(M)}\}$ \\
$\mathcal{A}$                 & Arm set, $\mathcal{A}$ = $\{a^{(1)}, a^{(2)}, \cdots, a^{(K)} \}$ \\
$\mathcal{I}$                 & Item set, $\mathcal{I}$ = $\{i^{(1)}, i^{(2)}, \cdots, i^{(N)}\}$ \\
$\mathcal{H}$                 & The hierarchy tree for item set partition.                             \\
$Ch(n)$                       & The set of child nodes of node $n$   \\
$Pa(n)$                       & The parent node of node $n$  \\
$I(n)$                        & The set of items mounted on node $n$  \\
$\mathcal{V}_u(t)$            & The receptive field of user $u$ at the $t$-th round \\
$\bm{X}_a$                    & The (static) embedding features of arm $a$, $\bm{X}_a \in \mathbb{R}^{d\times1}$     \\
$\eta$                        & The Gaussian noise of reward, $\eta \sim \mathcal{N}(0, \sigma^{2})$ \\
$i_{\pi}(t)$                  & The selected arm by policy $\pi$ at the $t$-th round                 \\
$r_{\pi}(t)$                  & Reward of policy $\pi$ at the $t$-th round                       \\
$\bm{\theta}_u$, $\bm{\theta}_u^{(l)}$               & Learnable parameter of user $u$. A superscript indicates the user parameter is for arms at level $l$ on $\mathcal{H}$.  $\bm{\theta}_u$ , $\bm{\theta}_u^{(l)}$ $\in \mathbb{R}^{d\times1}$  \\
\hline
\end{tabular}
\label{notation}
\end{table}

\section{Preliminary and problem}
We start by introducing the multi-armed bandit algorithms and the motivations of this paper. For better readability, we summarize most of the notations used throughout the paper in Table~\ref{notation}.  
\subsection{UCB for Recommender Systems}
The recommender system is regarded as an agent, where there are $M$ users and $N$ items. At each round $t = 1, 2, \cdots, T$ of interactions, given a user $u$, the agent recommends an item $i_{\pi}(t)$ to the user according a policy $\pi$. Then the agent receives a feedback $r_{\pi}(t)$ from the user, for example, if the user clicks on the item $i_{\pi}(t)$, $r_{\pi}(t)$ is 1 and otherwise it is 0. The optimal policy is denoted by $\pi^{*}$. The goal is to learn a good policy $\pi$, so that the cumulative regret over $T$ rounds, which is defined as below, is minimized:
\begin{equation}
    \bm{R}(T) = \sum_{t=1}^T (\bm{E}[r_{\pi^{*}}(t)] - \bm{E}[r_{\pi}(t)])
    \label{regret_eq}
\end{equation}
In practice, due to the absence of the optimal policy $\pi^{*}$, we maximize the cumulative reward $\sum_{t=1}^T \bm{E}[r_{\pi}(t)]$ instead, because maximizing cumulative reward equals to minimizing cumulative regret~\cite{li2016collaborative,wang2018hierarchical,yang2020exploring}.  

At the core of bandit algorithms is to find an optimal trade-off between exploitation~(to recommend fully based on user profiles learned from user interaction history) and exploration~(find out the new items which user may potentially love better), so that users diverse new interests have a certain chance to expose, meanwhile the system won't waste too many resources on items that users are not interested in.
Let's consider the (user-centric) LinUCB~\cite{li2010contextual} algorithm. Each item is regarded as an arm. At $t$-th round, when receiving a user visit request, 
the agent selects an arm $a_{\pi}(t)$ by:
\begin{equation}
    a_{\pi}(t) = \argmax_{a\in\mathcal{A}_t} R_a(t) + C_a(t)
    \label{linucb}
\end{equation}
The policy $\pi$ of LinUCB is a linear function between the feature vector $\bm{x}_{a}$ and user hidden parameter $\theta_u$, where the estimated reward is $R_a(t) = \bm{\theta}_{u}^T\bm{x}_{a} + {\eta}$, $\eta$ is a Gaussian random variable representing environmental noise, whose mean is zero and variance is $\sigma^2 \leq 1$, The upper bound $C_a(t)$ measures the uncertainty of the reward estimation.
The key point lies in how to determine the parameter $\bm{\theta}_u$ and the upper bound $C_a(t)$. With LinUCB, we have:
\begin{equation}
    \bm{\theta}_{u} = {\left({\bm{D}_t}^T \bm{D}_t + {I}_d\right)}^{-1}{\bm{D}_t}^T \bm{r}_{t}  
\end{equation}
\begin{equation}
    C_a(t) = \alpha \sqrt{\bm{x}_{a}^T (\bm{D}_t^T \bm{D}_t+\bm{I}_d)^{-1}\bm{x}_{a}}
\end{equation}
where $\bm{D}_t \in \mathbb{R}^{t \times d}$ is the matrix of interacted arms' features up to time $t$, $\alpha$ is a hyper-parameter to control the probability that the bound $C_a(t)$ holds, $\bm{r}_t \in \mathbb{R}^t$ is the user response vector up to time $t$.

\subsection{The Challenges}
However, as revealed in Eq.\eqref{linucb}, LinUCB needs to enumerate and calculate the score for every arm and then select the best one. In a modern recommender system, the number of items is usually very large (millions or even billions), which makes it impossible to calculate scores for all item. Thus, in the research community, a typical setting for existing literature is to randomly sample a small number $K$ (such as 50) arms from the entire $N$ arms at time $t$, and perform LinUCB on this small arm set $\mathcal{A}_t$; in industry, the bandit algorithms can only be applied to scenarios whose candidate pool is small, such as post-ranking stage of a recommender system, homepage most popular item ranking, ad creative ranking, etc. We argue that in order to fully explore users' potential interest, it is better to place the bandit module in the item retrieval stage (aka the recall stage) of a recommender system, where the candidate pool is the entire item set. Otherwise, in the post-ranking stage of a recommender system, the candidates are actually proposed by recommendation models and are strongly related to users' past behaviors. Thus, it is less meaningful to explore users' interest in the latter stages of a recommender system. To fully alleviate the closed-loop effect, in this paper, we advocate to explore users' interest in the entire space of item repository. However, to the best of our knowledge, there is no work studying how to make the bandit algorithm like LinUCB fit for a large candidate set.

To address the challenge, we propose to use a tree structure to partition the entire item space into multiple sub-spaces and build the hierarchical dependencies among items, to accelerate the exploration. Formally, we define the Framework~\ref{p0}:

\begin{Framework}\label{p0}
\textbf{Tree-based Exploration} The entire item set can be organized as a hierarchical tree structure $\mathcal{H}$, where nodes are linked to a subset of items that share some common topics or user interests, and nodes moving from top to bottom reflects the topics/interests partition being coarse-to-fine. During the tree-based exploration, we will first select a node according to some mechanism, then select an item from the candidates linked to this node. The user feedback on the selected item will not only update the item-wise user preference estimation, but also update the node-wise user preference estimation along the hierarchical path.
\end{Framework}
\section{Methodology}\label{method}
\subsection{Tree Structure Construction} \label{treebuild}
The tree structure plays a significant role in designing hierarchical bandit algorithms. Item category taxonomy can serve as the hierarchy tree. However, due to the imbalanced number of items under different category and lack of leveraging of users' collective behaviors, simply using the category taxonomy may lead to suboptimal performance, which is verified in Section \ref{experiment}.  In view of this, we first learn item embeddings based on item content and user co-click behaviors, then design a bottom-up clustering method based on K-Means clustering algorithm~\cite{likas2003global} to 
form a hierarchy tree for modeling dependencies among items. 


Specifically, to construct a tree structure with $L$ levels, at first, $N$ items are clustered into $k_L$ different subsets based on the similarity of item embeddings. We treat each subset as a new node on the tree, with an embedding vector being the average of all item embeddings belonging to this node. 
Afterward, these $k_L$ nodes will be further clustered into $k_{L-1}$ different subsets using K-Means and each subset will be treated as a new node on the tree, forming a parent-children relation. This step will be repeated several times until the depth of the tree structure researches $L$. As a result, the constructed tree structure, denoted by $\mathcal{H}$, contains $\{k_0, k_1, k_2, \cdots, k_L\}$ nodes at each level, where $k_0 = 1$ because only a root node appears at the first level. 
Intuitively, items within the same node are more similar to each other, thus the clustering results reflect the dependencies among items. In $\mathcal{H}$, only the root node does not have parent node, and leaf nodes have no children nodes.

\subsection{Hierarchical Contextual Bandit}
In this section, we introduce the proposed hierarchical contextual bandit ({\firstmodelname}) algorithm, which empowers a base bandit model to explore over the entire space of item repository. Our algorithm can be generalized to different bandit models, without loss of generality, we take LinUCB as the base model to explain the algorithm for clarity.

There are two types of arms: nodes on the hierarchy tree $\mathcal{H}$ and items mounted to the leaf nodes.
Each node on $\mathcal{H}$ represents a certain group of items. The feature vector of a leaf node is the average pooling of items mounted to it, and a non-leaf node's feature vector is the average pooling of its children nodes' feature vectors. The {\firstmodelname} algorithm makes decisions sequentially, starting from the root node to a leaf node. At any non-leaf node $n^{(l)}(t)$ at $l$-th level, the policy $\pi$ selects one of the child nodes from $Ch(n^{(l)}(t))$ by assuming the expected reward of an arm is linear in its feature vector, which is ${\bm{\theta}_{u}^{(l)}}^T\bm{X}_{n}$, and ${\bm{\theta}_{u}^{(l)}}$ is the latent parameters of a given user $u$ towards the nodes at level $l$,  ${\bm{D}^{(l)}} \in \mathrm{R}^{m \times d}$ is the matrix comprised of interacted items at $l$-th level, each row of ${\bm{D}^{(l)}}$ represents an item's feature vector. Applying ridge regression to the training samples to estimate the coefficients, we have:
\begin{equation}
    \bm{\theta}_{u}^{(l)} = {\left({\bm{D}^{(l)}}^T \bm{D}^{(l)} + \bm{I}\right)}^{-1}{\bm{D}^{(l)}}^T \bm{r}^{l} 
\end{equation}
Where $\bm{I}$ is an identity matrix and $\bm{r}^{l}$ is the vector of historical rewards at node level $l$. LinUCB also considers confidence interval to better estimate the arm payoff. Let ${\bm{A}^{(l)}} = {\bm{D}^{(l)}}^T{\bm{D}^{(l)}} + \bm{I}$. According to \cite{walsh2009exploring}, with probability $1-\delta$, the upper bound is:
\begin{equation}
    \left|{\bm{\theta}_{u}^{(l)}}^T\bm{X}_{n} - \bm{E}[r_{\pi^*}|\bm{X}_{n}]\right|\leq \alpha \sqrt{\bm{X}_{n}^T {\bm{A}^{(l)}}^{-1} \bm{X}_{n}}
\end{equation}
for any $\delta > 0$ and $\alpha = 1 + \sqrt{\ln(2/\delta)/2}$. In this way, the LinUCB algorithm tends to select an arm with:
\begin{equation}
\begin{aligned}
    n^{(l+1)}(t) = \argmax_{n\in Ch(n^{(l)}(t))} \left({\bm{\theta}_{u}^{(l)}}^T\bm{X}_{n} + \alpha \sqrt{\bm{X}_{n}^T {\bm{A}^{(l)}}^{-1} \bm{X}_{n}}\right)
\end{aligned}
\label{levelwiselinucb}
\end{equation}

If policy $\pi$ recommends $i_{\pi}(t)$ to a given user and receives the reward $r_{\pi}(t)$, similar as ~\cite{zhu2018learning}, then each node on $Path(root \to n^{(L)}(t))$ also receives the same reward $r_{\pi}(t)$. Therefore, the rewards of all selected nodes can be obtained, we can update the learnable parameters $\{\bm{\theta}_{u}^{(0)}$, $\bm{\theta}_{u}^{(1)}$, $\bm{\theta}_{u}^{(2)}$, $\cdots$, $\bm{\theta}_{u}^{(L)} \}$ at each level (where $\bm{\theta}_{u}^{(0)}$ means the parameter towards item arms, the other $\bm{\theta}_{u}^{(*)}$ means parameters towards node arms), which can be formulated as:
\begin{equation}
\begin{aligned}
    &\bm{A}^{(l)} \leftarrow \bm{A}^{(l)} + \bm{X}^{(l)}{\bm{X}^{(l)}}^T \\
    &\bm{b}^{(l)} \leftarrow \bm{b}^{(l)} + r_{\pi}(t)\bm{X}^{(l)} \\
    &\bm{\theta}_u^{(l)} \leftarrow {\bm{A}^{(l)}}^{-1}\bm{b}^{(l)}
\end{aligned}
\label{updatelinucb}
\end{equation}
Where $\bm{A}^{(l)}$ and $\bm{b}^{(l)}$ are initialized as $d$-dimensional identity matrix and zero vector respectively. $\bm{X}^{(l)}$ is the contextual embedding of the selected node at $l$-th level.

\begin{algorithm}[tbp]
\LinesNumbered
\KwIn{The tree structure $\mathcal{H}$ with depth as $L$, the total number of rounds $T$, hyper-parameter $\alpha$.}
\KwOut{The policy $\pi$.}
Initialize parameters of $\pi$;\\
\For{t = 1, 2, $\cdots$ ,T}{
receive a user $u$; \\
set current node as the root of $\mathcal{H}$; \\
\For{l = 1, 2, $\cdots$, L-1}{
    select a child node from current node with Eq.~\eqref{levelwiselinucb}; \\
    set current node as the selected child node;\\
}
select an item from the set of items of current~(leaf) node to user $u$ with Eq.~\eqref{linucb};\\
receive the reward from user $u$; \\
propagate rewards to all nodes in the path;\\
according to Eq.~\eqref{updatelinucb}, update parameters of $\pi$; \\
}
\caption{The pseudo-code of {\firstmodelname} algorithm}
\label{firstalg}
\end{algorithm}

\begin{figure}
    \centering
    \includegraphics[scale=0.4,trim=50 120 450 100,clip]{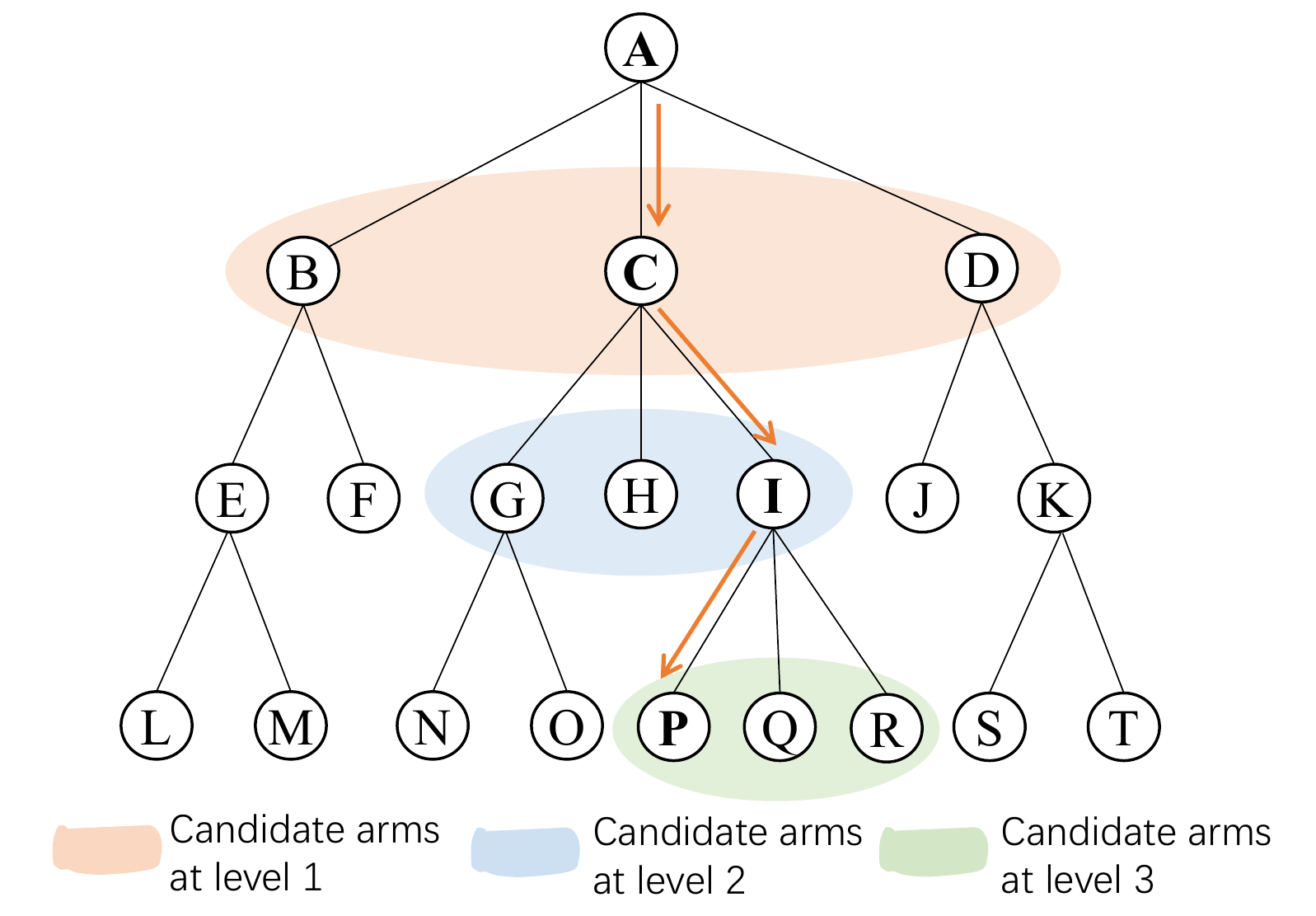}
    \caption{An illustration of \firstmodelname. The policy selects a path \{ A, C, I, P \} from root to a certain leaf node.}
    \label{linucb_dfs}
\end{figure}

The pseudo-code of {\firstmodelname} is provided in Algorithm~\ref{firstalg}.
To illustrate {\firstmodelname}, we offer a toy example shown in Figure~\ref{linucb_dfs}. It has three layers in the hierarchy tree. The agent makes three decisions sequentially, and finally select the path \{ $A, C, I, P$ \}. Then the agent will launch another bandit selection among the items mounted to the leaf node $P$. The reward on the selected item will impact the parameter estimation on the hierarchy tree $\{\bm{\theta}_{u}^{(0)}$, $\bm{\theta}_{u}^{(1)}$, $\bm{\theta}_{u}^{(2)}$, $\bm{\theta}_{u}^{(3)} \}$, by updating the reward history $\bm{r}^{(*)}$ and interaction history ${\bm{D}^{(*)}}$.

\subsection{Progressive Hierarchical Contextual Bandit}
The {\firstmodelname} learns the interests of each user via a sequential decision-making processes and always select the item from the arriving leaf node, which may lead to two problems: (1) the decisions made in upper levels severely impact the scope of lower-level nodes. Once the policy makes a bad decision at a certain level, the rest selections are all sub-optimal. The issue is especially true when the tree hierarchy is deeper. We call this phenomenon \textsl{error propagation}; (2) Users may be interested in more than one child node, thus the greedy selection may fail to capture the comprehensive interests of users. Therefore, we further propose a progressive hierarchical contextual bandit ({\secondmodelname}) algorithm for exploration in another manner on the tree.  
The main idea is that the policy continuously expands the receptive field from top to bottom according to the feedback obtained from historical exploration. We first give a definition of receptive field as follows.

\begin{Definition} \label{d0}
\textbf{Receptive field} is a personalized set of nodes representing the current potential interests for each user to explore. At the first round, the receptive field only consists of the root node (or is set with prior knowledge). With the exploration process progressing, the receptive field will be expanded (and reduced) when predetermined conditions are met in an adaptive top-down manner. The nodes in the receptive field are called visible nodes.
\end{Definition}

In {\firstmodelname}, only the leaf node is associated with a set of items. In contrast, in {\secondmodelname} we allow the policy to select a non-leaf node and then recommend an item from the item set associated with the non-leaf node. Hence, we have the Definition.~\ref{p3} to define the item set of each non-leaf node. 

\begin{Definition}\label{p3}
Given a non-leaf node $n$ and the set of child nodes $Ch(n)$, the item set of node $n$ will be the union of item sets of the nodes in $Ch(n)$, that is $I(n) = I(n_c^{(1)}) \cup I(n_c^{(2)})\cup \cdots \cup I(n_c^{(k)})$ and $Ch(n) = \{n_c^{(1)}, n_c^{(2)}, \cdots, n_c^{(k)}\}$.
\end{Definition}

At $t$-th round, the agent faces a user $u$ whose receptive field is denoted as $\mathcal{V}_u(t)$.  {\secondmodelname} algorithm treats each node in $\mathcal{V}_u(t)$ as an arm, and selects the arm~(denoted as $n(t)$) with highest estimated reward according to Eq.~\eqref{linucb}. Then another LinUCB algorithm is used to select one item $i_{\pi}(t)$ from the selected $I(n(t))$ and collect the feedback from the user. {\secondmodelname} directly selects an arm from the receptive field without performing sequential decision-making processes, which avoids the aforementioned concerns of {\firstmodelname}. 

\begin{figure}
    \centering
    \includegraphics[scale=0.4,trim=90 130 500 100,clip]{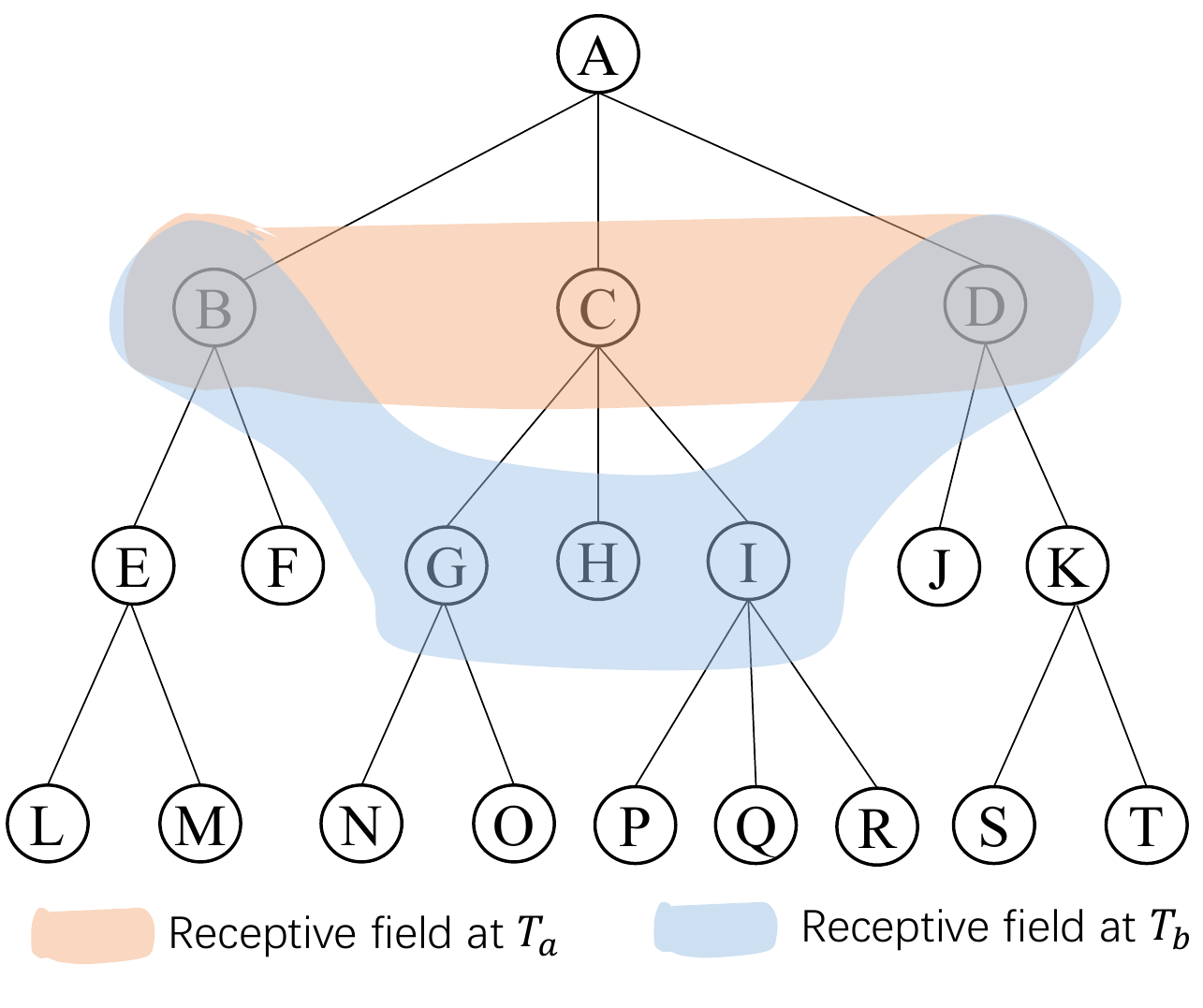}
    \caption{An illustration of {\secondmodelname}. At round $T_a$, the receptive field consists of nodes B, C and D; After several trials, at round $T_b$, node C meets the conditions of expansion, so the receptive field changes to nodes B, D, G, H and I}
    \label{linucb_bfs}
\end{figure}

Here we offer an example in Figure~\ref{linucb_bfs} for illustrating the expanding process. Assuming at round $T_a$, the receptive field of the user $u$ consists of three nodes: $B$, $C$ and $D$. In the next several rounds, if node $C$ is selected multiple times and received several positive rewards, making it meet the conditions of expansion, its children nodes $G, H, I$ will then be added into the receptive field to replace $C$. As a result, at round $T_b$, the receptive field includes nodes $B, D, G, H,I$. In this way, {\secondmodelname} expands the receptive field from coarse to fine and gradually discovers the  interests of users. 

A critical mechanism of {\secondmodelname} is how to expand the receptive field. 
Since the tree nodes are organized in different granularity, the nodes at top levels represent the coarse interests of users while the nodes at bottom levels depict specific interests of users. We want the receptive field be able to quickly converge to the leaf nodes, thus we set the expansion conditions as follows: for a non-leaf node $n$ at the $l$-th level of $\mathcal{H}$, if (1) it has been selected at least $\lfloor q\dot\log{l} \rfloor$ times, and meanwhile (2) the average reward on this node is larger than $p\dot\log{l}$~($0 \leq p \leq1$), then we expand the receptive filed by replacing it with its children. $q$ and $p$ are hyper-parameters. The $\log{l}$ means that the nodes at a top-level are easier to be expanded than those at a low level. One can also design more flexible rules for expansion according to the actual application scenario. Overall, the pseudo-code of the {\secondmodelname} is available in Algorithm~\ref{secondalg}.

\begin{algorithm}[tbp]
\LinesNumbered
\KwIn{The tree structure $\mathcal{H}$ with depth as $L$, the total number of rounds $T$, the original receptive field $\mathcal{V}(0)$, hyper-parameters $\alpha$, $p$ and $q$.}
\KwOut{The policy $\pi$.}
Initialize parameters of $\pi$;\\
\For{t = 1, 2, $\cdots$ ,T}{
receive a user $u$; \\
select a node $n$ from current receptive field $\mathcal{V}_u(t)$; \\
select an item from $I(n)$ to user $u$ with Eq.~\eqref{linucb};\\
receive the reward from user $u$; \\
\If{node $n$ satisfies the expansion conditions}{
    add the nodes in $Ch(n)$ into the receptive field to replace node $n$;
    }
update parameters of $\pi$; \\
}
\caption{The pseudo-code of {\secondmodelname} algorithm}
\label{secondalg}
\end{algorithm}

\subsection{Regret Analysis}
As defined in Eq.~\eqref{regret_eq}, the regret is defined as the difference between the expected reward under hindsight knowledge and the actual reward under the algorithm. The regret bound we would like to obtain is established on a premise that the clustering structure is known to the algorithm ahead of time, which is consistent with the scenario of this paper. In this case, each cluster is viewed as an independent arm, according to ~\cite{auer2002using,abbasi2011improved,chu2011contextual}, the regret bound is up to logarithmic terms $\ln(T)$, $\ln(N)$, and $\ln(1/\delta)$. By hiding the logarithmic factors with notation $\widetilde{\mathcal{O}}$, the cumulative regret over $T$ rounds is bounded with probability $1-\delta$ as:

\begin{equation}
    R(T) = \widetilde{\mathcal{O}}\left( \sum_{k=1}^{K} (\sigma d + ||\bm{X}_k||\sqrt{d})\sqrt{T} \right)
    \label{regret01}
\end{equation}

In Eq.~\eqref{regret01}, we shall assume that $||\bm{X}_k|| = 1$ for all clusters~\footnote{Without loss of generality, we can conduct $l_2$ normalization over the contextual vector associated with each arm.}. Then as proven by ~\cite{gentile2014online}, one can replace $\sqrt{T}$ of each arm by a term formulated as $\sqrt{T}(\frac{1}{K} + \sqrt{\frac{|V_k|}{N}})$, where $|V_k|$ is the size of $k$-th cluster and $N$ is the total number of items.  As a result, the regret bound becomes:

\begin{equation}
    R(T) = \widetilde{\mathcal{O}}\left( (\sigma d + \sqrt{d})\sqrt{T} \left(1+\sum_{k=1}^K\sqrt{\frac{|V_k|}{N}}\right) \right)
    \label{regret02}
\end{equation}

In Eq.~\eqref{regret02}, according to ~\cite{gentile2014online}, the worst case occurs when each cluster has the same size $\frac{N}{K}$, leading to the regret bound:

\begin{equation}
    R(T) = \widetilde{\mathcal{O}}\left( (\sigma d + \sqrt{d})\sqrt{KT} \right)
    \label{regret03}
\end{equation}

Here we first discuss the regret bound for \firstmodelname. For simplicity, we assume that the number of clusters are reduced $m$ times that of the previous level. At the beginning, each item is treated as a cluster, i.e. the number of clusters should be $N$. In this case, the regret bound holds $\widetilde{\mathcal{O}}\left( (\sigma d + \sqrt{d})\log_{m}(N)\sqrt{mT} \right)$. For \secondmodelname, the receptive field expands in a progressive manner, assuming the final size of receptive field is $r$, the regret bound is at most $\widetilde{\mathcal{O}}\left( (\sigma d + \sqrt{d})\sqrt{rT} \right)$.  As proven in ~\cite{chu2011contextual}, if the arm set is fixed over time and contains $N$ arms, the regret bound of the contextual bandits with linear payoff functions is up to $\mathcal{O}(\sqrt{Td}{\ln}^{3/2}(NT\ln(T)/\delta)))$. It is significantly larger than the regret bound of {\firstmodelname} and {\secondmodelname} due to the higher order of logarithmic terms and $d \ll N$, $m \ll N$ and $r \ll N$ in most instances. Therefore, our proposed algorithms can improve the exploration efficiency substantially by reducing the cumulative regret. Moreover, the exploration time complexity is reduced from $\mathcal{O}(N)$ to $\mathcal{O}(\log(N))$ with the help of hierarchy.

\section{Experiments} \label{experiment}

\subsection{Experimental Settings}
\subsubsection{\textbf{Datasets}}
We perform experiments on two public recommendation benchmark datasets, basic statistics are shown in Table~\ref{dataset}.

\squishlist
\item \textbf{MIND}\footnote{\url{https://msnews.github.io/index.html}}~\cite{wu2020mind}: The MIND dataset is the largest public benchmark dataset for news recommendations so far, which is constructed from the click logs of Microsoft News.  We use Sentence BERT~\cite{reimers2019sentence} to train news embeddings from their contents, and adopt a GRU~\cite{okura2017embedding} as the user model to fine-tune news embeddings from the sequence of click logs. 
\item \textbf{Taobao}\footnote{\url{https://tianchi.aliyun.com/dataset/dataDetail?dataId=649&userId=1}}: The Taobao dataset is constructed from user behaviors of Taobao for E-commerce recommendations. Similar to MIND dataset, we also utilize GRU to learn item embeddings from the sequence of user behavior logs. 
\squishend

\begin{table}[tbp]
\small
\topcaption{Overview of Datasets}
\setlength{\tabcolsep}{0.5mm}
\renewcommand\arraystretch{1.0}
\begin{tabular}{c|c|c|c|c}
\hline
\textbf{Dataset} & \textbf{\#users} & \textbf{\#items} & \textbf{\# categories} & \textbf{\# interactions} \\ \hline \hline
MIND             & 1,000,000        & 161,013          & 285                     & 24,155,470               \\
Taobao           & 987,994          & 4,162,024        & 9,439                  & 100,150,807              \\ \hline
\end{tabular}
\label{dataset}
\vspace{-6mm}
\end{table}
\subsubsection{\textbf{Baselines}}
We compare the proposed algorithms against the following related and competitive bandit algorithms:
\squishlist
\item \textbf{LinUCB}~\cite{li2010contextual} is a classical contextual bandit algorithm. It only works with item-level recommendation.
\item \textbf{HMAB}~\cite{wang2018hierarchical} organizes arms into hierarchy tree purely by domain knowledge.  It utilizes category information to model the dependencies among items. Then the algorithm selects a path from root node to a leaf node, and the leaf node is an item.
\item \textbf{ICTRUCB}~\cite{wang2018online} formulates the item dependencies as the clusters on arms. Different from our methods, it does not consider the hierarchy.
\item \textbf{ConUCB}~\cite{zhang2020conversational} utilizes key-terms to organize items into different subsets to represent dependencies among items. The algorithm occasional conversation with users and leverages conversational feedbacks on key-terms from users to accelerate the speed of bandit learning.
\squishend

\subsubsection{\textbf{Our Methods and Variants}}
Our goal is to propose a generic algorithm that can empower different bandit models to be more effective on large-scale item set exploration. Therefore, our main experiments contain two groups: first, with LinUCB as the base bandit model, to compare our algorithms (i.e., HUCB and pHUCB) with the aforementioned baselines (because most of the listed baselines are based on LinUCB); second, with three different base bandit models, including LinUCB, Thompson Sampling (TS)~\cite{agrawal2013thompson}, and $\epsilon$-greedy, to verify whether our algorithms are effective under different settings. In the second group of experiments, we also compare two variants of our models:
\squishlist
\item \textbf{CB-Category}: It borrows the idea from HMAB~\cite{wang2018hierarchical} by using the prior knowledge, i.e. the category taxonomy, to assign items into different subsets. Each subset will be treated as an arm, the policy first selects a subset and then recommends an item from the subset to users with a base bandit algorithm.  
\item \textbf{CB-Leaf}: It is a variant of ICTRUCB~\cite{wang2018online}. Since the ICTRUCB is designed for context-free bandits, to keep a fair comparison, we also utilize K-Means clustering on item embeddings to assign items into different subsets. Here, we treat the leaf nodes of $\mathcal{H}$ as the clustering results. Each leaf node will be treated as an arm, the policy first selects a leaf node and then recommends an item from the node to users with a base bandit algorithm. 
\squishend

Here, \textsl{CB-} can take a value from \{ LinUCB, Thompson Sampling (TS), and $\epsilon$-greedy \}, our experiments are separated into three groups. For example, if the \textsl{CB-} model is LinUCB, then the involved variants are \textsl{LinUCB-Category}, \textsl{LinUCB-Leaf}, and our final models are \textsl{HUCB} and \textsl{pHUCB}. For all models, including our proposed ones and baselines, we set the maximum times of score-computing per round to 50 for a fair comparison. For example, in LinUCB, if the number of arm candidates 
is 1000, then originally we need to compute 1000 scores per round to select the best one in estimation, which exceeds our budget, so we will randomly sample 50 arms from the 1000 arms and then perform the LinUCB on the small set. For hierarchical CB methods, the budget is evenly distributed to each level, e.g., for pHCB, we have two levels of bandit, then each level will get a budget of 25.

\subsubsection{\textbf{Evaluation}} We evaluate the performance of different bandit algorithms with \textit{off-policy user simulator evaluation}. To reduce the biases of the simulation, we utilize the IPS estimator~\cite{schnabel2016recommendations,huang2020keeping,saito2020large}, which is a standard method used for \textit{off-policy evaluation} of bandit algorithms~\footnote{\url{https://groups.google.com/g/open-bandit-project}}. IPS learns to re-weigh the training samples by the propensity score to learn an unbiased simulator. The simulator is trained on the whole data to make the best use of information. This evaluation enables us to compare the performance of candidate hypothetical policies without expensive online A/B tests. Specifically, the simulator learns the unbiased embeddings of users and items, and the reward of a user $u$ towards an item $i$ is derived from the inner product of their embeddings. 

\subsubsection{\textbf{Reproducibility}} For MIND dataset, each item is represented by a 64-dimensional embedding vector. The dataset has been split for training/validation/test, only the click logs of the training set is used for learning item embeddings and building tree structures. The users without history logs or impression logs are removed. The click is treated as positive feedback and non-click is treated as negative feedback. The hierarchy tree structure is set to \{ 1, 100, 10000 \}, which means there is 10000 leaf nodes, and only one layer of non-leaf nodes.
For Taobao dataset, each item is represented by a 32-dimensional embedding vector. As the same method does in ~\cite{zhu2018learning}, we remove the users who have less than 10 behaviors or the items appearing less than 10 times. We randomly select 10000 users for testing and 1000 users for validation, and the behavior logs of the rest users are used for learning item embeddings. The click is treated as positive feedback and the negative feedback is generated via negativing sampling.  The hierarchy tree structure is set to \{ 1, 50, 5000, 50000 \}, so that number of items mounted to a leaf node is less than 100.

For the LinUCB-based algorithms, all learnable parameters are initialized as all zero matrices or vectors, and the hyper-parameter $\alpha$ is set as 0.5. The Gaussian prior is used to design Thompson Sampling-based algorithms for contextual bandit. For $\epsilon$-greedy-based algorithms, the $\epsilon$ is set as 0.05. With the help of validation set, the hyper-parameters $q$ and $p$ of {\secondmodelname} and its variants are set as 10 and 0.1 respectively. Follow the setting in \cite{zhang2020conversational}, we consider a general configuration in which at each round, the computational cost is limited as 50. Note here the arms can be nodes or items depending on different bandit algorithms. The learning rate for training GRU is 0.001, the hidden size is the same as the embedding size, and optimizing with Adam optimizer~\cite{kingma2014adam}. All algorithms are implemented with Python and PyTorch, and repeated 10 times to report the average performance. 
The code and processed datasets will be released after acceptance of the paper for easy reproducibility.

\begin{table}[htbp]
\small
\setlength{\tabcolsep}{1.8mm}
\renewcommand\arraystretch{1.2}
\topcaption{Cumulative rewards of different algorithms}
\label{cr_tab}
\begin{tabular}{c|ccc|ccc}
\hline
\textbf{Dataset} & \multicolumn{3}{c|}{\textbf{MIND}}           & \multicolumn{3}{c}{\textbf{TaoBao}}         \\ \hline
\textbf{Round}   & \textbf{100} & \textbf{1000} & \textbf{2000} & \textbf{100} & \textbf{1000} & \textbf{2000} \\ \hline \hline
LinUCB           & 4.76         & 158.41        & 357.91        &  10.70       &  106.12       & 212.26              \\
HMAB             & 4.41         & 217.40        & 520.97        &  10.74       &  120.89       & 255.22               \\
ICTRUCB          & 5.60         & 300.83        & 709.05        &  10.98       &  135.99       & 290.33               \\
ConUCB           & 7.47        & 188.50       & 409.23       & 13.04       & 273.86       & 584.84       \\ \hline
HUCB             & 7.48        & 407.82       & \textbf{918.38} & 16.20    & 320.68       & 699.29       \\
pHUCB            & \textbf{7.93} & \textbf{419.74} & 866.60 & \textbf{16.51} & \textbf{391.36}  & \textbf{806.66}        \\ \hline
\end{tabular}
\end{table}

\begin{figure*}[t]
    \centering
    \subfigure[MIND, LinUCB]{
    \includegraphics[scale=0.125]{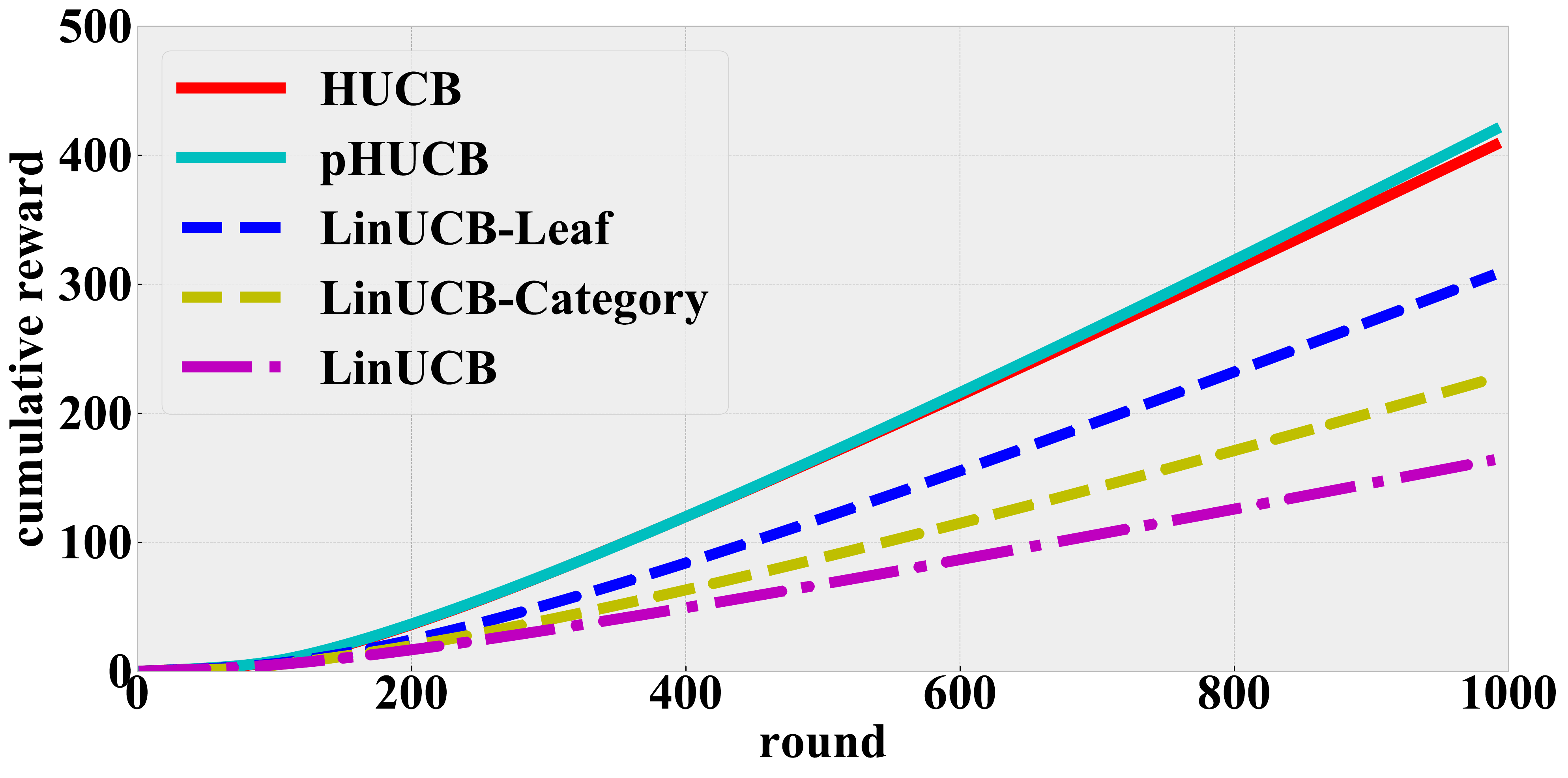}
    }
     \subfigure[MIND, Thompson Sampling]{
    \includegraphics[scale=0.125]{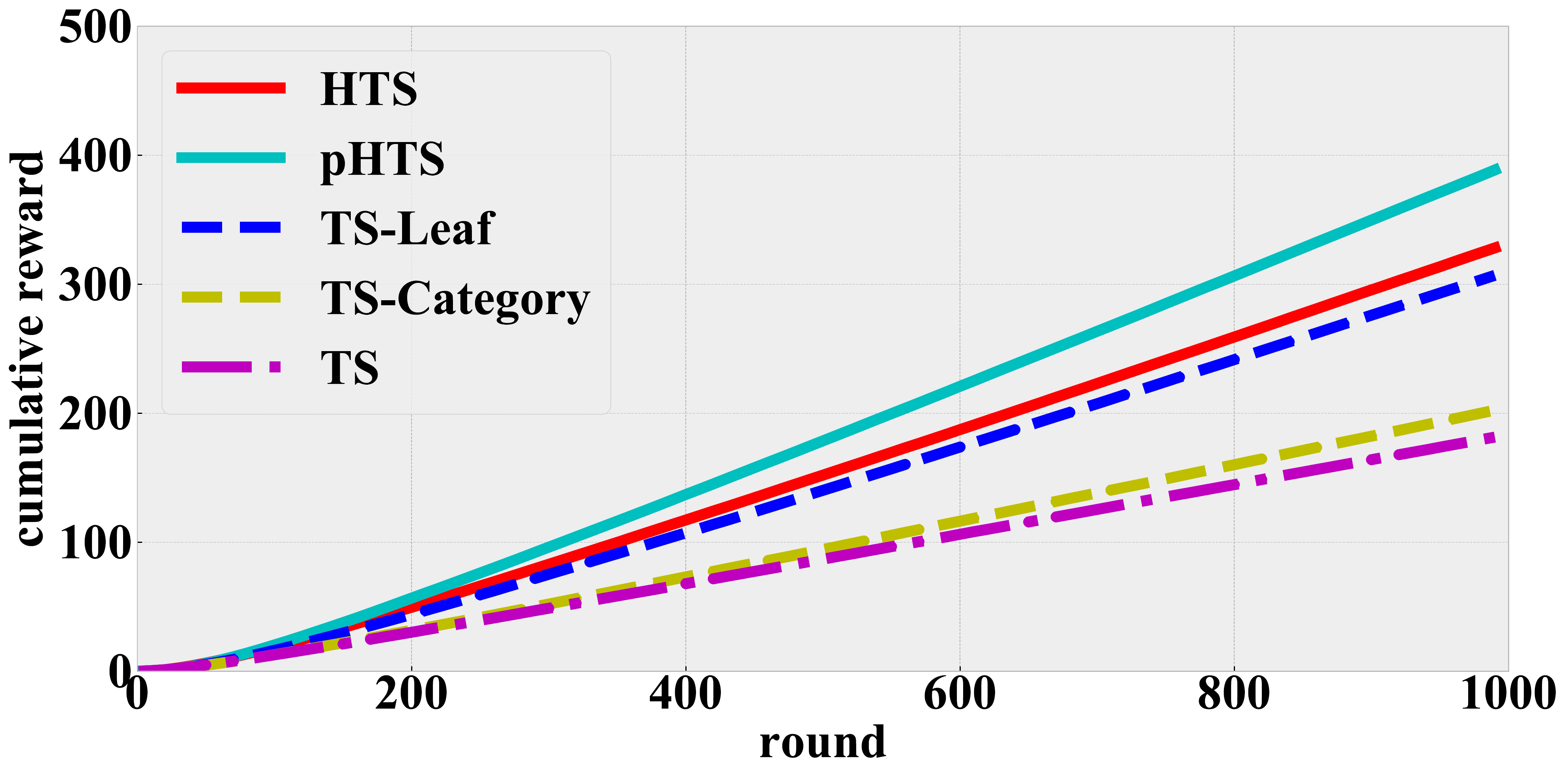}
    }
     \subfigure[MIND, $\epsilon$-greedy]{
    \includegraphics[scale=0.125]{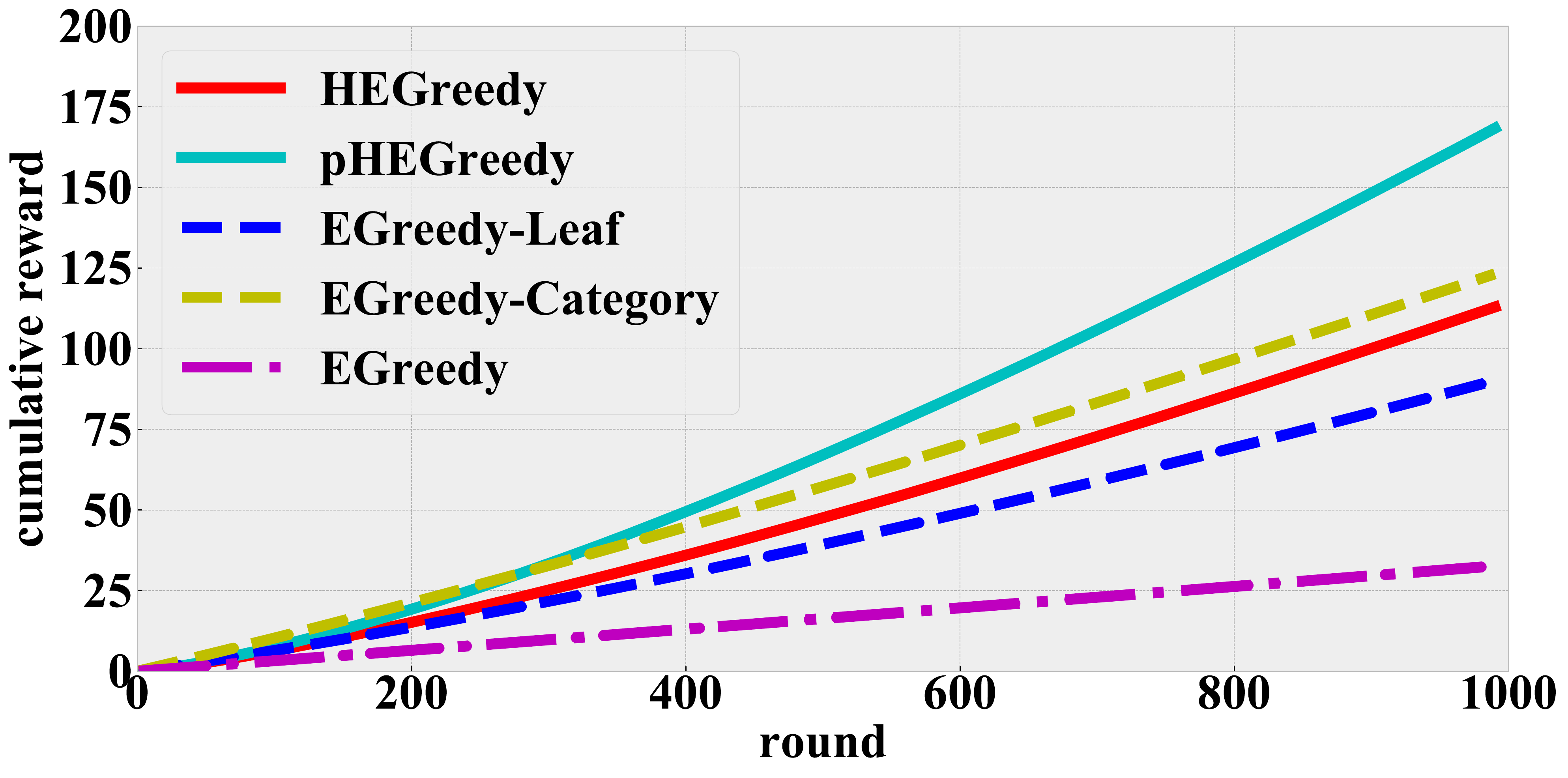}
    }
    \subfigure[Taobao, LinUCB]{
    \includegraphics[scale=0.125]{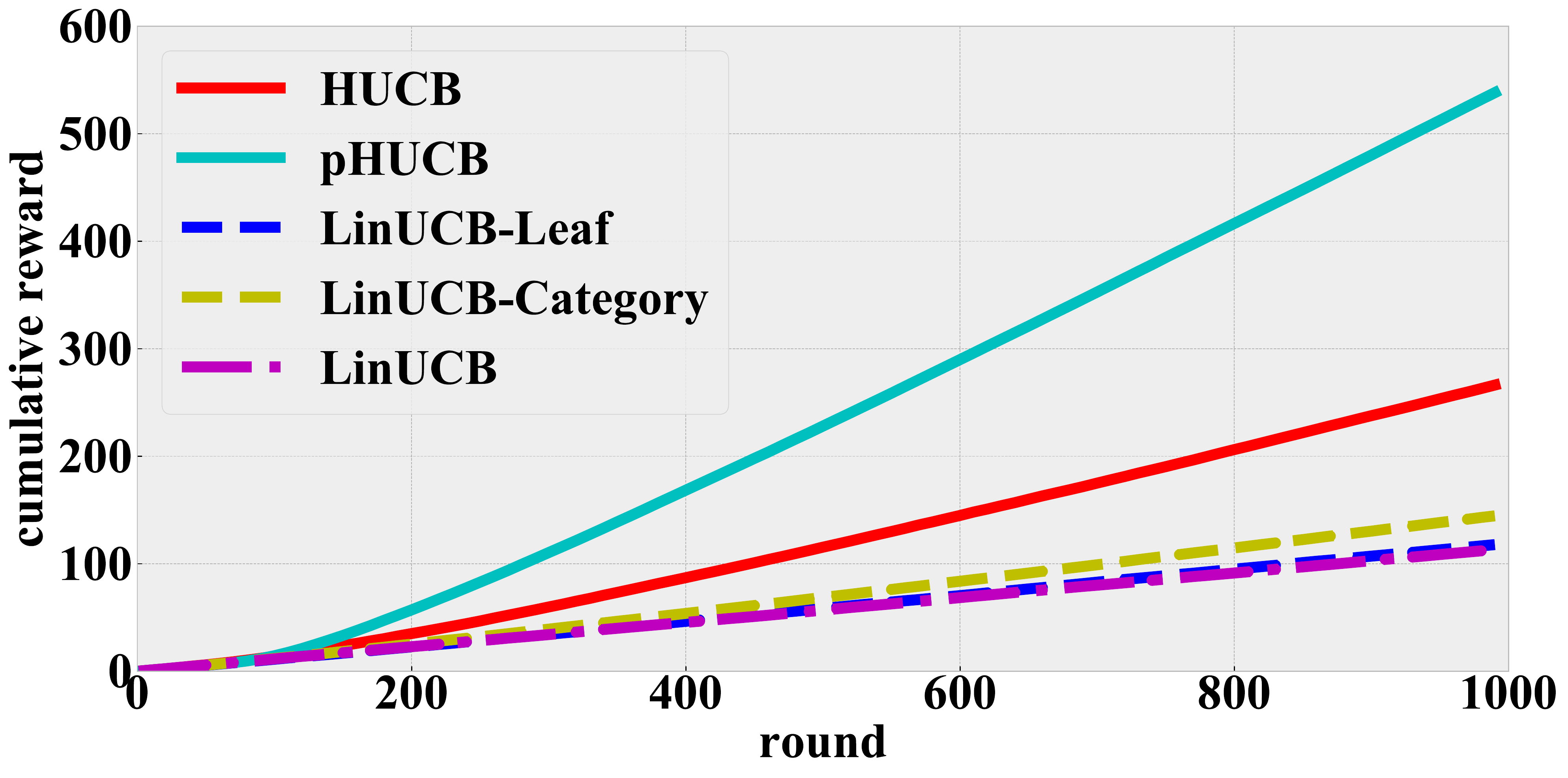}
    }
     \subfigure[Taobao, Thompson Sampling]{
    \includegraphics[scale=0.125]{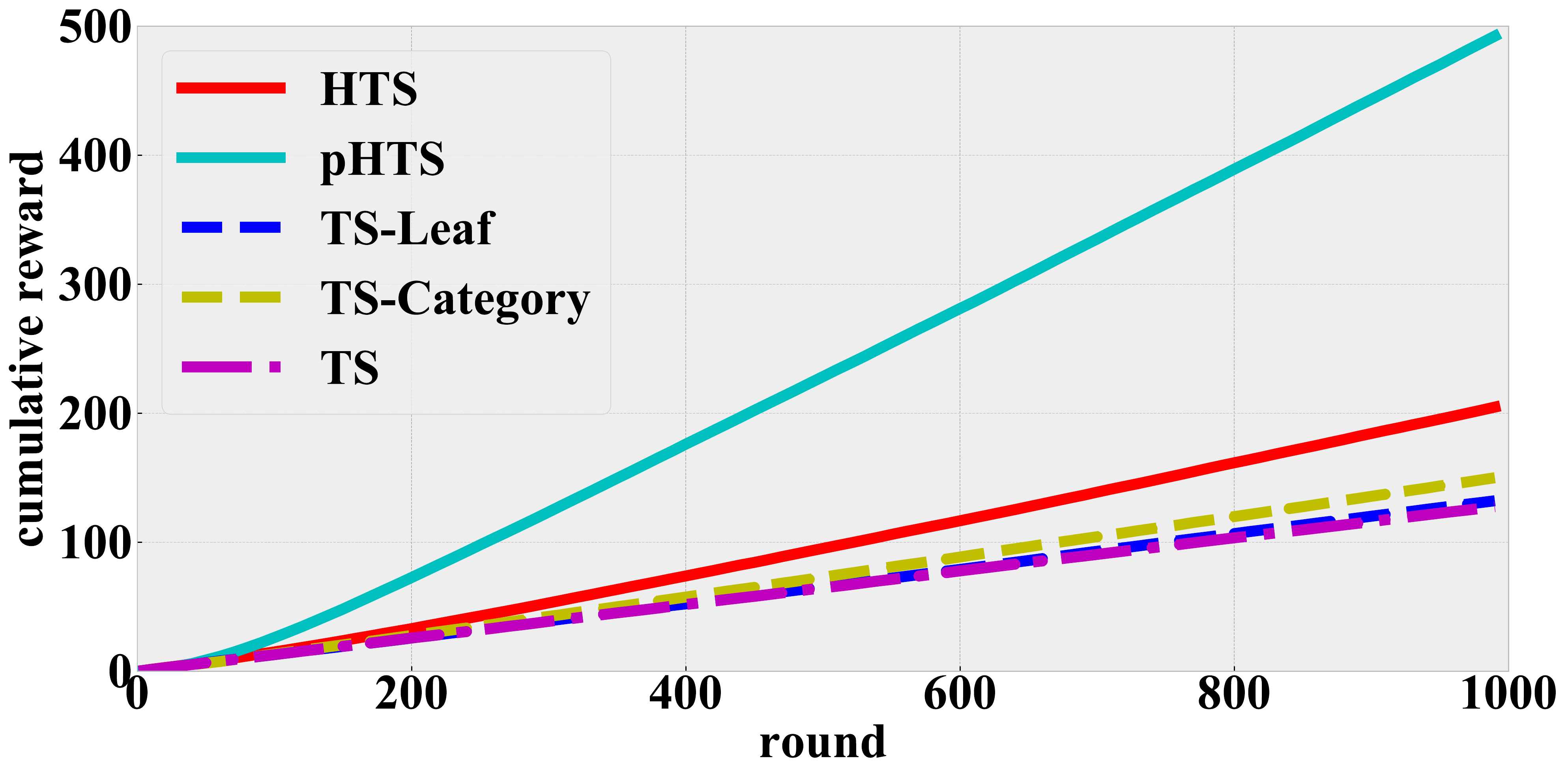}
    }
     \subfigure[Taobao, $\epsilon$-greedy]{
    \includegraphics[scale=0.12]{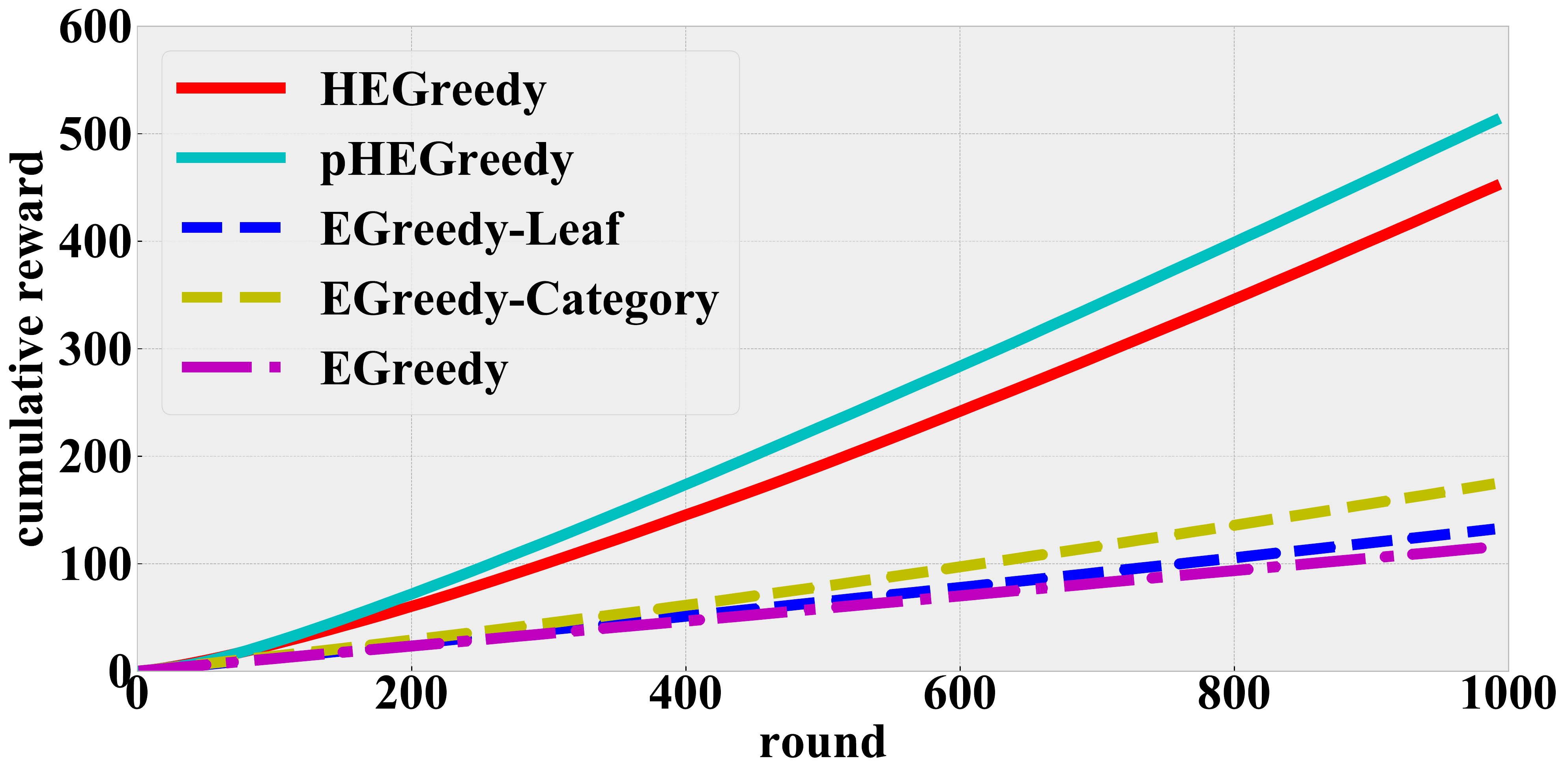}
    }
    \caption{Cumulative rewards of our algorithms and variants based on LinUCB, Thompson Sampling and $\epsilon$-greedy, on the MIND dataset and Taobao dataset, respectively.}
    \label{flexibility_exps}
\end{figure*}

\subsection{Experimental Results}

\subsubsection{\textbf{Comparison with Baselines}}
Since all baseline algorithms are on the basis of LinUCB, we also choose LinUCB as the base algorithm for HCB and pHCB to keep a fair comparison. Note that HCB and pHCB can work with different base algorithms, and we discuss their generality in sec.~\ref{general}. We compare the cumulative rewards over 100/1000/2000 rounds\footnote{Considering that the two datasets have different number of users and items, for better alignment, here we denote \textsl{round} as \textsl{one pass of all users receiving one recommended item}.} of different algorithms in Table~\ref{cr_tab}, and the best results are presented in bold. As can be seen, our proposed algorithms, HUCB and pHUCB, outperform all baselines across different datasets consistently at different rounds, and pHUCB is generally better than HCB. For example, on the largest TaoBao dataset, At 100/1000/2000 rounds, the performance of pHUCB was improved by 54.3\%, 268.7\%, and 280\% compared with LinUCB, respectively. Although HMAB, ICTRUCB, and ConUCB achieve higher cumulative rewards over LinUCB, there is still a considerable gap between these algorithms and ours. The superiority of pHUCB over HUCB further verifies that by expanding the receptive field in a progressive manner, the pHCB algorithm is able to better discover the comprehensive interests of users.

\subsubsection{\textbf{Flexibility and Variants Study}} \label{general}
Next, we report the cumulative reward of our algorithms and their variants, in Figure~\ref{flexibility_exps}, based on three different base bandit algorithms. From the experimental results, we have the following observations. 
\squishlist
\item Constructing item dependencies in the form of clusters indeed helps a lot in accelerating the exploration. This can be verified from that both our algorihtms (HCB and pHCB) and their variants (CB-category and CB-Leaf) outperform the corresponding base bandit model.
\item HCB and pHCB achieve the highest cumulative rewards on both two datasets in most of the cases, indicating that the proposed hierarchical algorithms are effective. As we can see, the performance of baseline algorithms has a noticeable gap between our proposed algorithms. This result is reasonable since our proposed methods introduce the hierarchy knowledge to take the item dependencies into consideration, which greatly improves the efficiency of exploration. Although the two variants, such as CB-Category and CB-Leaf, also organize items into different clusters, they fail to model the coarse-to-fine item dependencies as tree structures. Apart from that, the pHCB algorithm beats other methods, which verifies that the progressive exploration can adaptively discover the diverse interests of users with a receptive field. 
\item Our algorithms have strong flexibility. We have tested the performance with base models varying in \{ LinUCB, Thompson Sampling, $\epsilon$-greedy \}, the conclusions are consistent, which proves our proposed frameworks can well generalize to various bandit algorithms.  

\squishend

\subsubsection{\textbf{Parameter Sensitivity}}
In this subsection, we study the impacts of hyper-parameters. Since the pHCB performs best in most of the cases, we particularly study the key hyper-parameter of it, i.e., $q$ and $p$, which control the expansion conditions: $q$ determines the number of trails at one arm and $p$ indicates the threshold of average reward for expanding child nodes. To study their impacts, we take pHUCB as an example and vary $q$ from $\{0,5,10,15,20,25\}$ and $p$ from $\{0,0.05,0.1,0.15,0.2,0.25\}$ to see how the final rewards (at round 1000) will be affected.

\begin{figure}[tb]
    \centering
    \subfigure[MIND]{
    \includegraphics[scale=0.13]{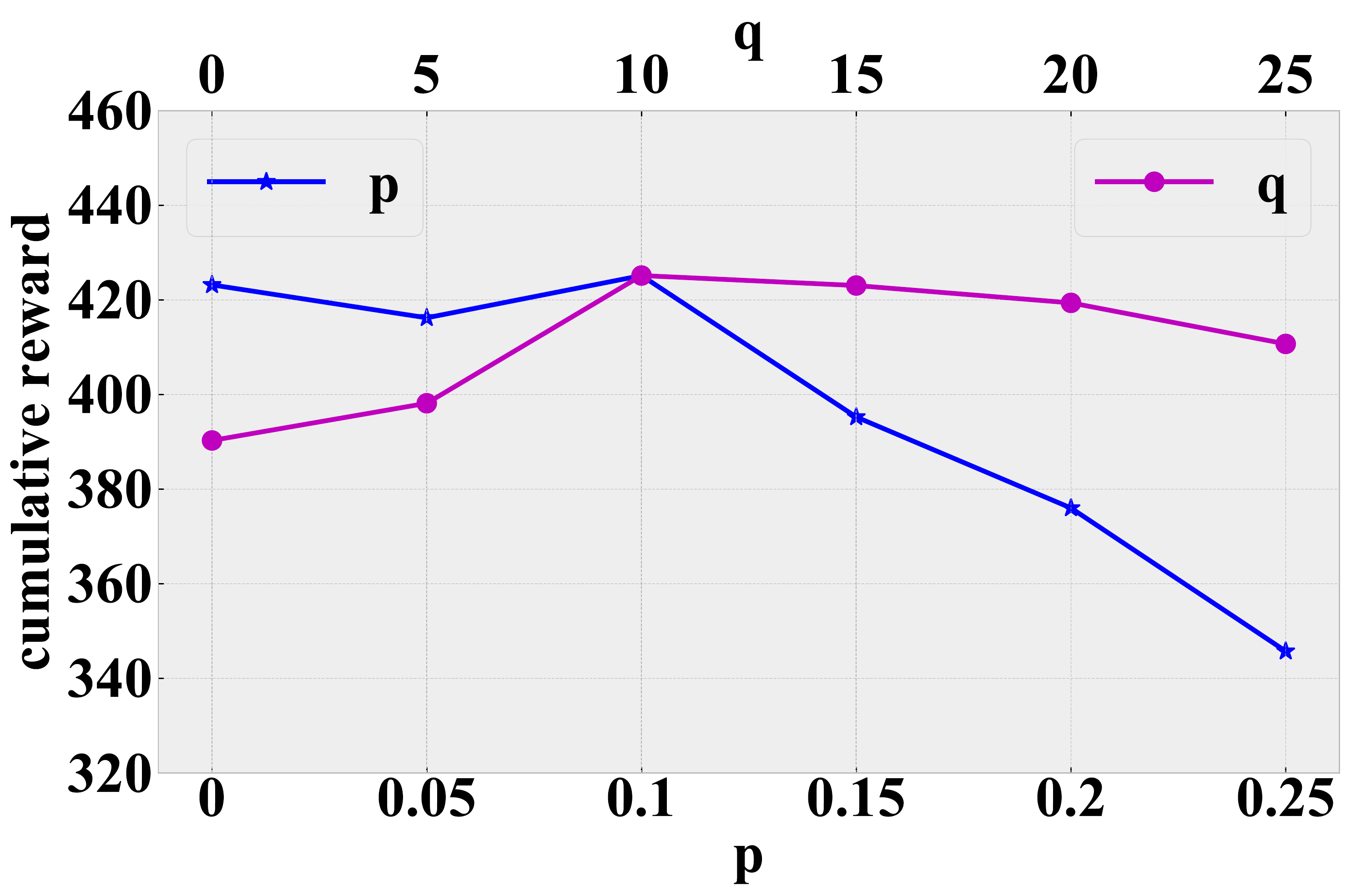} 
    }
    \subfigure[Taobao]{
    \includegraphics[scale=0.13]{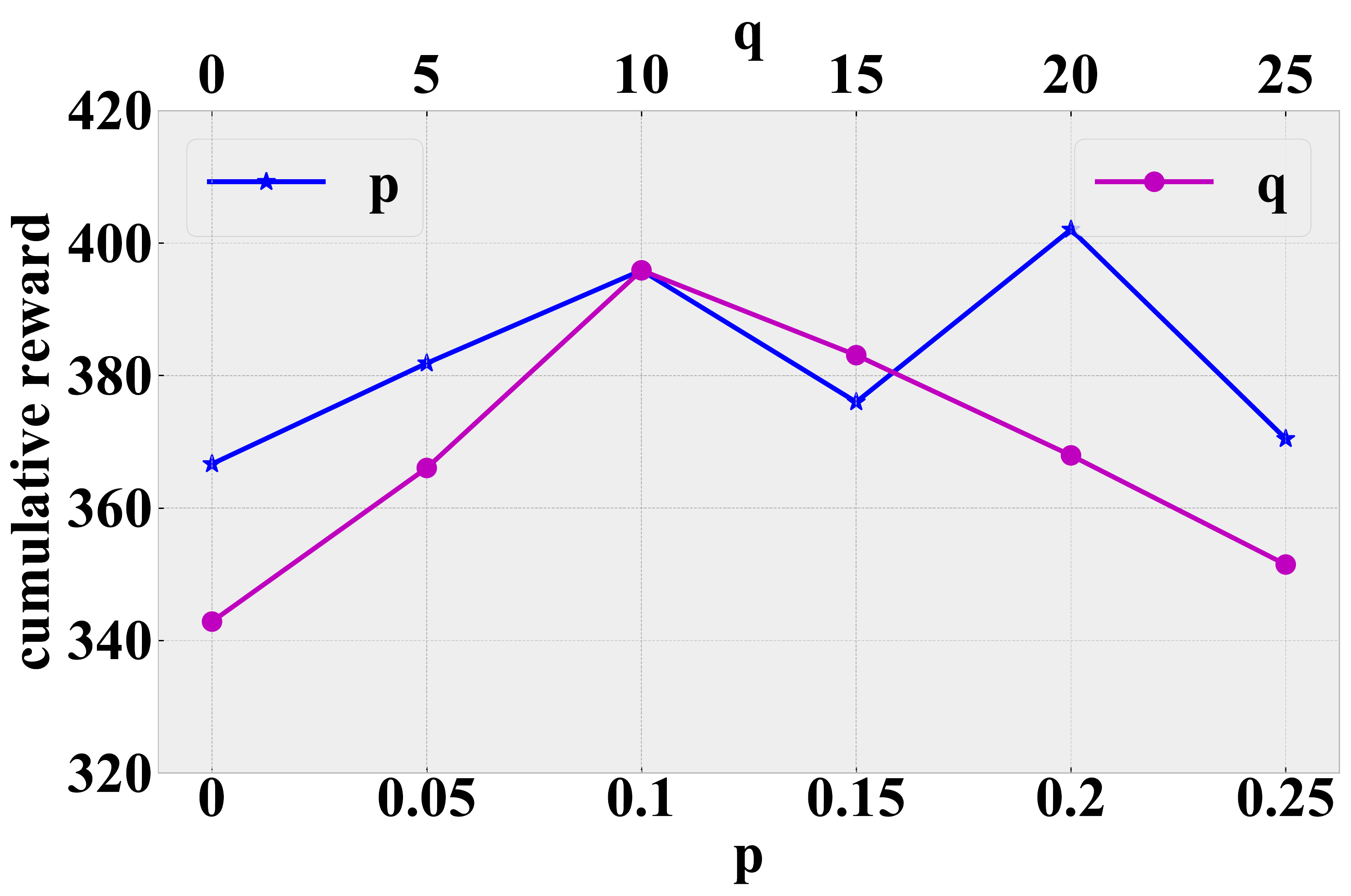}
    }
    \vspace{-4mm}
    \caption{Effect of hyper-parameters of pHUCB.}
    \label{phucb_pq}
    \vspace{-6mm}
\end{figure}

As shown in Figure~\ref{phucb_pq}, different hyper-parameters have a noticeable influence on the cumulative reward of pHUCB algorithm over 1000 rounds. For MIND dataset, if $q$ is too small or $p$ is too large, the cumulative rewards become worse since the former makes the expansion conditions unreliable and the latter makes the receptive field difficult to expand. As for the Taobao dataset, the trend of impacts caused by $q$ is similar to the MIND dataset. Meanwhile, the model is less sensitive with parameter $p$. Overall, from Figure~\ref{phucb_pq} we learn that a suggested configuration is $p=0.1$ and $q=10$.

\subsubsection{\textbf{Alleviate Closed-Loop Effects}} \label{scratch training} 
Typically, recommender models trained from historical logs are designed for exploitation purposes, here we called them exploitation models. Such recommendation systems often suffer from closed-loop effects~\cite{jadidinejad2020using} because they only learn users' interests from historical logs, but they do not have the ability to explore new interests of users. In contrast, the contextual bandit algorithms are more effective to break the information cocoons with exploration strategies. In order to verify that our model is able to explore the potential interests of users thus alleviate the closed-loop effects, we design an additional experiment as follows.

We select three exploitation models, i.e., Linear model, GRU, and Transformers~\cite{sun2019bert4rec} as baselines. These exploitation models are first pre-trained by the historical logs of existing users to get the deployed models. Then, we use exploitation models as well as our proposed models as "deployed models" to serve users. Specifically, in this stage, for each new user (whose logs are not used in the first pre-trained stage), we randomly sample only three clicked items as visible historical logs for generating her initial user embedding with a deployed model. Then we can recommend two hundred items to the user with a deployed model and collect the user's feedback. Note here the user embedding will be refreshed once the model receives positive feedback. This stage is performed for every deployed model respectively. The third stage is about evaluating the quality of impression logs produced by the deployed models. 
We utilize the collected historical logs together with all the rest historical logs of existing users  (which are used in the first stage) as training samples to train an evaluating model~(here we use the matrix factorization (MF) model as the evaluating model, each user and item will be mapped to an embedding vector. MF-based collaborative filtering method is one of the most popular models for personalized recommendations) and evaluate the trained model on the same test samples for a fair comparison. 
In order to prove the advantages of our bandit algorithms in exploring user interests, we select two hundred of test users with the most diversified interests as new users: we calculate the Gini impurity~\cite{loh2011classification} of the historical items clicked by the user according to the category of items. Obviously, the larger the Gini impurity, the more diverse the interests of the user. The users of the training set are treated as existing users.

\begin{table}[tb]
\small
\setlength{\tabcolsep}{1.2mm}
\renewcommand\arraystretch{1.1}
\topcaption{Test LogLoss and AUC of different algorithms}
\label{loss_auc}
\begin{tabular}{c|cc|cc}
\hline
\textbf{Dataset} & \multicolumn{2}{c|}{\textbf{MIND}} & \multicolumn{2}{c}{\textbf{TaoBao}} \\ \hline
\textbf{Method}  & \textbf{LogLoss}     & \textbf{AUC}          & \textbf{LogLoss}    & \textbf{AUC}   \\ \hline\hline
Linear           & 1.679$_{\pm 0.005}$  & 0.703$_{\pm 0.005}$   & 0.693$_{\pm 0.001}$ & 0.530$_{\pm 0.001}$                \\
GRU              & 1.759$_{\pm 0.004}$  & 0.686$_{\pm 0.003}$   & 0.688$_{\pm 0.001}$ & 0.535$_{\pm 0.002}$               \\
Transformer      & 1.377$_{\pm 0.008}$  & 0.695$_{\pm 0.006}$   & 0.683$_{\pm 0.001}$ & 0.546$_{\pm 0.001}$               \\ \hline
HUCB             & \bf{0.681}$_{\pm 0.004}$  & \bf{0.720}$_{\pm 0.003}$   & \bf{0.660}$_{\pm 0.001}$ & \bf{0.649}$_{\pm 0.002}$  \\
pHUCB            & \bf{0.680}$_{\pm 0.005}$  & \bf{0.723}$_{\pm 0.002}$   & \bf{0.661}$_{\pm 0.002}$ & \bf{0.647}$_{\pm 0.003}$                \\ \hline
\end{tabular}
\end{table}

We report the test log loss~(LogLoss) and area under curve~(AUC) score in Table~\ref{loss_auc}. 
We can observe that both our methods, HUCB and pHUCB, achieve much higher performance than exploitation models including the Linear model, GRU, and Transformers, which demonstrates that our proposed models can effectively help to alleviate the closed-loop effects in recommender systems.

\section{Conclusion} \label{conclusion}
In this paper, we propose a general hierarchical bandit framework for entire space user interest exploration.
Specifically, we design two algorithms, i.e., {\firstmodelname} and {\secondmodelname}. The {\firstmodelname} algorithm makes a sequence of decision-making tasks to find a path from the root to a leaf node, while the {\secondmodelname} progressively expands the receptive field in a top-down manner to explore the user interests, which is more flexible and also achieves more satisfactory results. Extensive experiments are conducted to demonstrate the effectiveness of the proposed framework on two real-world datasets with three different base bandit algorithms. In the future, we plan to combine our methods with the start-of-the-art deep learning methods to estimate reward for making more reasonable decisions. Moreover, we assume the items are static in this paper by fixing the tree structure unchanged. It would be interesting to extend the proposed frameworks to the non-static setting, which has not been well studied yet.

\newpage
\bibliographystyle{ACM-Reference-Format}
\bibliography{ref}

\end{document}